\documentstyle[aps,twocolumn]{revtex}

\newcommand{\beq}{\begin{equation}}
\newcommand{\eeq}{\end{equation}}
\newcommand{\beqa}{\begin{eqnarray}}
\newcommand{\eeqa}{\end{eqnarray}}

\def\<{\langle}
\def\>{\rangle}

\def\opone{\leavevmode\hbox{\small1\kern-3.8pt\normalsize1}}
\def\I{{\cal I }}

\begin{document}
\title{Quantum cryptography with fewer random numbers}

\author{N. Gisin}
\address{\small\em Group of Applied Physics, University of Geneva, 1211
Geneva 4, Switzerland}
\date{\today}

\maketitle

\begin{abstract}
We prove that in the BB84 quantum cryptography protocol Alice and Bob do not need to make
random bases-choice for each qubit: they can keep the same bases for entire blocks of
qubits. It suffices that the raw key consists of many such qubit-blocks. The practical
advantage of reducing the need for random number is emphasized.
\end{abstract}
\vspace{1cm}


The security of quantum cryptography \cite{BB84,GisinRMP} is based on two main ingredients. The first
one is well known: the celebrated {\it no cloning theorem} \cite{NoCloning}, or - 
equivalently - Heisenberg uncertainty relations\cite{BourennanePRL}, or {\it entanglement-without-signaling}
\cite{GisinSignalingCloning}. The second ingredient is also an absolute necessity, although often only
mentioned implicitly: truly random choices on both sides. Clearly, quantum cryptography should
use quantum randomness, i.e. the only physical source of true randomness. But, in practice this
is a severe constrain, because a complete protocol requires a huge amount of random numbers,
from Alice's state choices to Bob's basis choices and for the random choices and random permutations needed for
error correction and privacy amplification.

In this note we prove that the rush for quantum random numbers can be reduced during the 
quantum communication phase of the protocol without impairing on the security against
individual attacks.
The idea is the following: Alice and Bob use the same basis for large blocks
of qubits (hence, here individual attacks refers to attacks block-by-block). 
During sifting, if they happen to have chosen different bases, the entire block is
disregarded. For each block Alice and Bob make new and independent random choices of bases.
In the limit of large blocks, this reduces the rate at which random bits are needed 
by about one half \cite{halfreduction}. At first sight, one may think that this makes Eve's life easier:
she knows that all the qubits within a block are coded in the same basis. But we shall 
see that this is not the case: Eve's optimal attack provides her with no more Shannon
information, for a given error rate QBER, than in the usual case where Alice and Bob make random base-choices
for each qubit. Consequently, provided the raw key consists of (infinitely) many blocks of qubits,
the Csisz\'{a}r, and J. K\"orner theorem \cite{CsiszarKorner} applies: if the mutual 
information Alice-Bob is larger than either the
mutual informations Eve-Alice or Eve-Bob, then Alice and Bob can distil a secret key.
In practice the mentioned many blocks should be processed together for error correction and privacy
amplification; hence the block size should not be too large, if not error correction and privacy 
amplification consumes too much time.

The proof that Eve can not take advantage of the fact that all qubits within a block are
coded in the same basis (unknown to Eve) follows an argument given by Xiang-Bin \cite{XiangCohEve}.
It runs by contradiction. Assume that, for a given QBER, Eve can extract 
an averaged information \I 
per qubit of a block of length $n\ge2$. Then, Eve can extract at least as much information
attacking each qubit one by one. For this she prepares $n-1$ pairs of qubits in the singlet
state. She keeps one qubit per singlet and uses the others to simulate a $n$ qubit block,
see Fig. 1.
Once Alice and Bob announced the basis used for this block, i.e. after sifting, Eve
measures the kept qubits in the announced basis, thus preparing effectively all the $n$ qubits
of the simulated block in the same basis, as illustrated in Fig. 1. Hence, she can extract \I information
per qubit from the simulated block, including \I information on the qubit sent by Alice.

This rather simple argument is of great practical value, assuming that the result also holds against
coherent attacks (i.e. attacks on the qubits of all blocks).
Indeed, in realistic implementations \cite{idQ} Alice sends out several millions of qubits per second,
and the trend is clearly towards even higher rates. This implies megabits of quantum random
numbers, a difficult though not impossible task. Reducing this factor by quasi one half,
without reducing the security, is clearly advantageous. For example, in the Plug-\&-Play 
configuration we do already send blocks of qubits from Alice to Bob, in order to circumvent
Rayleigh backscattering \cite{RibordyQCRayleigh}. 

The result presented in this note softens the 
random number generation bottleneck. 

\small
\section*{Acknowledgments}
Supported by the Swiss Center "Quantum Photonics", and by the Swiss OFES
within the frame of the European project RESQ.

\normalsize
\section*{Figure Captions}
Eve is assumed to have a quantum machine (represented by the unitary operator U) acting
on blocks of $n$ qubits plus $m$ ancillas.  After the unitary interaction Eve keeps her $m$ ancillas.
If Alice sends only one qubit (or if Eve likes
to attack the qubits one-by-one), Eve can simulate a n-qubit block by adding to Alice's
qubit the halfs of $n-1$ singlets.
After Alice and Bob announced the basis, Eve measures the other halves of the singlets in
this basis, thus effectively preparing the simulated block as $n$ qubits all coded in the
same basis.
\end{document}